## Highlights

### Comparison of Code Quality and Best Practices in IoT and non-IoT Software

Nour Khezemi,Sikandar Ejaz,Naouel Moha,Yann-Gaël Guéhéneuc

- Method for selecting equivalent 94 IoT and non-IoT systems software from GitHub.
- Comprehensive computation and analysis of various metrics and systems.
- An in-depth analysis of examples of IoT systems to illustrate how our code metrics values manifest in IoT system codebases.
- A revisited software engineering best practices list for IoT

# Comparison of Code Quality and Best Practices in IoT and non-IoT Software

Nour Khezemi[a,*,1], Sikandar Ejaz[b,*,2], Naouel Moha[a,3] and Yann-Gaël Guéhéneuc[b,4]

[a]*École de Technologie Supérieure, Montreal, Quebec, Canada*
[b]*Concordia University, Montreal, Quebec, Canada*

ABSTRACT

**Context**: IoT systems, networks of connected devices powered by software, require studying software quality for maintenance. Despite extensive studies on non-IoT software quality, research on IoT software quality is lacking. It is uncertain if IoT and non-IoT systems software are comparable, hindering the confident application of results and best practices gained on non-IoT systems.
**Objective**: Therefore, we compare the code quality of two equivalent sets of IoT and non-IoT systems to determine whether there are similarities and differences between the two kinds of software. We also collect and revisit software-engineering best practices in non-IoT contexts to apply them to IoT.
**Method**: We design and apply a systematic method to select two sets of 94 non-IoT and IoT systems software from GitHub with comparable characteristics. We compute quality metrics on the systems in these two sets and then analyse and compare the metric values. We analyse in depth and provide specific examples of IoT system's complexity and how it manifests in the codebases. After the comparison, We systematically select and present a list of best practices to address the observed difference between IoT and non-IoT code.
**Results**: Through a comparison of metrics, we conclude that software for IoT systems is more complex, coupled, larger, less maintainable, and cohesive than non-IoT systems. Several factors, such as integrating multiple hardware and software components and managing data communication between them, contribute to these differences. Considering these differences, we present a revisited best practices list with approaches, tools, or techniques for developing IoT systems. As example, applying modularity, and refactoring are best practices for lowering the complexity.
**Conclusion**: Based on our work, researchers can now make an informed decision using existing studies on the quality of non-IoT systems for IoT systems. Developers can use the list of best practices to minimise disparities in complexity, size, and cohesion and enhance maintainability and code readability.

## 1. Introduction

With the increasing use of IoT systems, it is vital to assess the quality of the source code of the software running (on) these systems [1]. These systems often operate in critical environments (e.g., healthcare, transport, and infrastructure management), and any flaw in the code could lead to failures, posing significant risks. In consequence, the quality of their source code significantly impacts their functionality, security, and reliability, making code assessment a critical component in their development lifecycle.

We believe comparing code quality between IoT and non-IoT software systems is essential for establishing the best system development and maintenance practices. IoT systems' unique constraints, such as limited resources and distributed architectures, necessitate a close analysis of code quality [2]. In smart cities, poor coding techniques in IoT systems might introduce various security flaws, including data exposition, integrity, and confidentiality. IoT devices work in dynamic situations, necessitating specific quality characteristics such as scalability and adaptability [3].

Many studies exist on the software quality of non-IoT systems, but there is a lack of research on the quality of IoT systems. In particular, we lack information about whether the software for IoT systems is comparable to non-IoT software. Without this knowledge, we cannot apply the results and best practices proven suitable for non-IoT systems with confidence to the IoT systems.

Developers can better understand the differences between IoT and non-IoT systems by doing a comparative analysis via metrics. They can adjust non-IoT best practices considering the unique requirements of IoT systems.

Existing research, while examining non-IoT and IoT system's software quality [4, 5, 6], primarily focuses on assessing non-IoT software code quality. This less attention on IoT systems creates a gap in understanding their software quality, raising a critical idea to compare the software quality of IoT and non-IoT systems. Without this comparison, we cannot confidently apply the results and best practices from non-IoT system software to the IoT system software because it lacks the necessary foundation and specificity. As a result, the efficacy and reliability of such practices are not guaranteed when applied to IoT system software. The comparison between IoT and non-IoT system software thus serves as a first step toward understanding the difference between the types of systems, tailoring existing practices according to

nour.khezemi.1@etsmtl.ca (N. Khezemi);
sikandar.ejaz@concordia.ca (S. Ejaz); Naouel.Moha@etsmtl.ca (N. Moha);
yann-gael.gueheneuc@concordia.ca (Y. Guéhéneuc)
ORCID(s): 0009-0001-6524-8539 (N. Khezemi); 0000-0001-7347-0765
(S. Ejaz); 0000-0001-9252-9937 (N. Moha); 0000-0002-4361-2563 (Y.
Guéhéneuc)

IoT system software specificities, and highlighting the need to develop new strategies suited for IoT system software.

Recognising the importance of this comparison, we developed a systematic methodology. This method allowed us to systematically collect, analyse, compare, and evaluate 94 comparable IoT and non-IoT systems. We provide a descriptive and in-depth analysis of the two types of systems. We present specific examples of IoT systems analysed in depth to illustrate how our code metrics values manifest in IoT system codebases.

Leveraging our findings, we propose an updated, systematically selected list of best practices to address the observed difference between the code of IoT and non-IoT systems.

This work employs a well-defined process with the following contributions:

1. Our first contribution is the method for selecting equivalent 94 IoT and non-IoT systems software from GitHub for ensuring the integrity and validity of our comparative analysis, minimising potential biases of our research. The selection process ensures that the chosen IoT and non-IoT systems are comparable regarding the number of stars and forks, forming a solid foundation for our evaluations.

2. Our second contribution stems from our comprehensive computation and analysis of various metrics and systems. Metrics are essential for our in-depth analysis, enabling a granular examination of the results based on detailed code evaluations. We provide an in-depth analysis of examples of IoT systems to illustrate how our code metrics values manifest in IoT system codebases.

3. Finally, we systematically select, discuss, and present a revisited software engineering best practices list for IoT systems, selected from the literature for each category of code metrics we studied. We show that by tailoring best practices, including code optimization techniques, modularity, and the use of design patterns, our study provides targeted solutions to address challenges such as high code complexity, low maintainability, and readability issues in IoT systems, offering a better understanding for the development of efficient and sustainable codebases.

Our comparison results highlight key differences between the two systems, such as complexity, cohesion, code size, and maintainability. We discuss their implications for IoT systems development. We found that developing software for IoT systems presents greater complexity than non-IoT systems, affecting the overall code quality. Considering these differences, we provide a revised list of best practices for developing IoT systems as a target solution. Our work demonstrates that future work is needed to implement the identified best practices list on IoT systems, and evaluating its effect is necessary to address issues such as complexity, size, and coupling.

The rest of this paper is organized as follows: Section 2 provides an overview and discussion of related work. Section 3 presents the research methodology of the study. Section 4 presents a quantitative analysis and discusses our comparison results. Section 5 presents an in-depth analysis of some IoT systems regarding the qualitative comparison results. Section 6 contains specific examples that illustrate the complexity of IoT systems and how it manifests in IoT codebases. Section 7 presents the practical implications of our comparison results for IoT systems development. Section 8 discusses results on best practices, while Section 9 presents threats to validity. Finally, Section 10 presents conclusions and future work directions.

## 2. Related Work

Klima *et al.* [7] summarised relevant code quality metrics from IoT systems and assessed their impact on general systems quality based on ISO/IEC standards.

They categorize those metrics into size (Lines of Code), complexity (Cyclomatic Complexity), coupling (Response For Class), etc. These metrics offer an accurate evaluation of IoT systems' code quality, and we will use and present them in our comparison, enabling us to systematically assess and juxtapose code quality between these IoT and non-IoT systems software.

While our study shares a similar approach in utilizing these established IoT systems code quality metrics, our focus extends beyond the evaluation of metrics. We undertake a comprehensive comparative analysis between IoT and non-IoT systems, leveraging these metrics to explore the nuanced differences and shared traits between these software paradigms.

Corno *et al.* [8] investigated open-source software development in IoT and non-IoT systems, analysing 60 projects. They found significant differences in development processes, developer specialisation, and code reusability between these two types of systems. Their study also examined developer contributions, file modifications, specialisation, and project maturity by analysing project dependencies.

Our study measures code metrics to compare IoT and non-IoT systems code quality. Although Corno *et al.*'s approach differs in research objectives and methods, these studies complement each other to understand IoT systems.

While previous literature hinted at the complexity of IoT systems [8, 9, 10, 11], we conducted a quantitative comparison to assess this complexity compared to non-IoT systems, which will be presented in this work.

Larrucea *et al.* [11] emphasised the lack of established software engineering best practices for IoT systems and highlighted the need for effective guidance in engineering IoT systems.

To address this gap, we select, study and provide a set of best practices for IoT system development from the literature. To the best of our knowledge, no prior work has compared the code quality of IoT and non-IoT systems. Our study aims to identify similarities or differences between IoT and non-IoT systems by analysing specific software metrics, offering a straightforward and effective approach for selecting comparable systems. Our work is pioneering the study of IoT system quality, advancing the understanding and enhancement of IoT development practices.

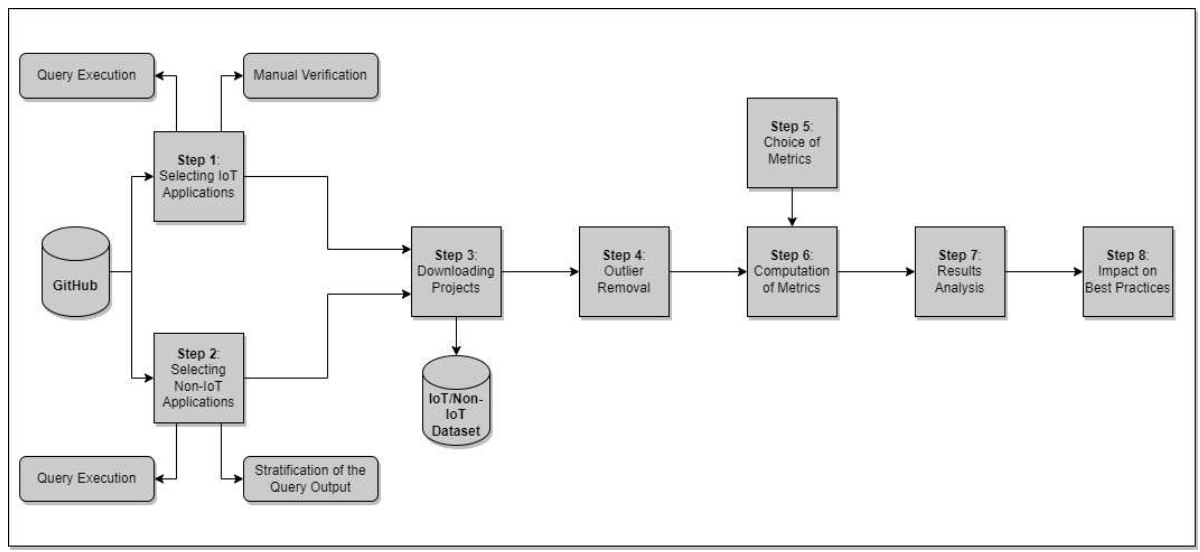

**Figure 1:** Methodology

## 3. Method

Finding comparable IoT and non-IoT systems is challenging, given the need for matching criteria like stars, forks, size, programming language, classes, and files. The approach of matching all those criteria did not yield an important number of systems on which we could base our comparison, so we adapted our selection process to focus on similar numbers of stars and forks. The popularity of a repository, as indicated by stars and forks, can reflect its relevance within a certain domain or for a particular use case [12]. GitHub stars offer users a means to convey their appreciation for repositories [8]. When two repositories share similar popularity, it implies they are valued within their respective categories. This approach ensures attention and recognition from the GitHub community.

Our methodology was influenced by Politowski *et al.* [13] and Corno *et al.* [8], with eight key steps as in Figure 1. In October 2022, we initiated Step 1 of our process by collecting popular IoT systems from GitHub. We started by filtering repositories based on topics, specifically focusing on those falling under 'IoT', 'Internet-of-things', 'Internet-of-things', 'EIoT', 'IIoT', and 'Internet of everything' or 'Industrial Internet of Things' topics. GitHub topics function as labels for categorising repositories according to their intended purpose, subject matter, or community [8].

After completing the filtration process, we successfully identified repositories associated with the IoT topics. To prioritise repositories with significant popularity and positive user evaluations, we sorted them based on the number of stars they had received, arranged in descending order.

In the final selection, we included the top repositories that were openly accessible with the highest number of stars. It is worth noting that a substantial portion of GitHub repositories does not pertain to software development. As a result, we conducted a manual inspection to exclude repositories unrelated to software (e.g., tutorials, documentation pages) and to apply some defined selection criteria that we will present later.

Moving on to Step 2, we studied and defined queries based on the criteria of the selected IoT systems. We employed execution query and stratification techniques to select comparable non-IoT projects.

In Step 3, we downloaded the selected projects into a database for further analysis.

In Step 4, we identified outliers through manual examination and subsequently eliminated them from the dataset.

In Step 5, we selected a comprehensive list of metrics that we intend to compute for IoT and non-IoT in Step 6.

In Step 7, we analysed and compared the computed metrics to derive meaningful insights and observations.

Finally, in Step 8, based on our findings of the comparison between IoT and non-IoT. We systematically gathered and reviewed practices for non-IoT to adapt them to the context of IoT.

We present our methodology by posing a series of questions that guided our selection process for IoT systems, non-IoT systems, metrics, and tools. In the following subsections, we systematically present each of them to motivate and explain our choices.

### 3.1. Which artifacts will we use to base our comparison?

We could use various software artifacts, including documentation, code, bug reports, chat logs, or execution logs [14] in the comparison process. We choose to focus on source code because it is the common basis for any software system describing its behaviour and functionality.

### 3.2. How Will We Compare the Two Sets of Systems?

We compare the two sets of systems using metrics because they offer a quantitative and objective way to assess quality. Metrics provide numerical values that allow for direct comparisons, reducing subjectivity and offering a clear basis for evaluating strengths and weaknesses.

We recognise while metrics are valuable, they may not provide a complete picture of system quality. To conduct a thorough assessment, robust quality models that consider various dimensions and factors are essential. Our work is an initial step in gathering vital insights by measuring and comparing quality metrics. This contributes to the development of more comprehensive quality models.

### 3.3. What Category of Metrics?

Code metrics are categorised by properties such as size, redundancy, complexity, coupling, unit test coverage, cohesion, code readability, security, code heterogeneity, and maintainability [7]. In our study, when statically comparing code, we selected metrics from different categories: size, complexity, cohesion, coupling, code readability, and maintainability. We did not explore security aspects, which have received extensive attention in both systems [15]. We chose to exclude unit test coverage and effectiveness categories because we focused on static code aspects. The redundancy category was also excluded as we consider it closely related to code readability and maintainability.

### 3.4. Which Tools We Use to Compute Metrics

There is a diversity of tools that we can use to compute those metrics. We opted for two tools due to their frequent use [16] and because they can measure the maximum of the list of metrics that we presented in Table 1. These tools provide comprehensive insights into system complexity, maintainability, and size, aligning perfectly with our research objectives.

Scitools Understand is designed to assist in understanding, evaluating, and verifying source code [17].

It supports a variety of languages and offers the possibility of measuring a variety of code metrics.

Multimetric is a Python library for creating multiple metrics [18]. It is designed to make it easy to build complex and multidimensional metrics that can be used in a variety of applications. The library provides a comprehensive set of APIs and utilities; we are using one of the APIs to measure our metrics. With Multimetric, we can quickly create, combine, and analyse multiple metrics in a single codebase.

In conclusion, Scitools Understand and Multimetric were selected due to their ability to handle multi-language support and provide a comprehensive analysis of a large list of metrics, ensuring accuracy in our measurements.

### 3.5. Which Metrics Are We Using?

There is an extensive list of code metrics under the categories presented from [7] in Table 1.

We selected well-known metrics that could be computed using the two tools chosen, Understand and Multimetric. Our focus was on choosing metrics that are not exclusively applicable to IoT systems but are more general, allowing us to effectively analyze and compare both types of systems.

| Category | Metric |
|---|---|
| Size | Lines Of Code (LOC), Estimated rebuild value (ERV), Unit Interface Size (UIS), Average Unit Size (US), Number of Not Architectural Components (NAC), Number of Classes and Files |
| Complexity | Cyclomatic Complexity (CC), Halstead Volume (HV), WMC-McCabe, Number of Children (NOC), Number of Thing Interconnections (NTI), Depth of Inheritance Tree (DIT) |
| Coupling | Response For Class (RFC), Coupling Between Objects (CBO), Number of Incoming calls per modules (INC) |
| Cohesion | Lack of Cohesion of Methods (LCOM), Conceptual Cohesion of Classes (C3), Ratio of Cohesive Interactions (NRCI) |
| Code Readability | Comment Percentage and Comment to Code Ratio (CP) |
| Maintainability | Maintainability Index (MI) |

**Table 1**
Metrics Categories

Table 2 presents the metrics that we compute using the chosen tools and their formulas. The motivation behind choosing these metrics lies in their collective ability to provide multifaceted insights into various aspects of code quality, ranging from system size and complexity to maintainability and readability. The selection aims to capture diverse dimensions that collectively contribute to software quality assessment.

### 3.6. Which Systems Are We Choosing for The Comparison?

We obtained the sets of systems from GitHub. GitHub provides a wide variety of applications that can be used to gain insights into software development trends, project management, and best practices. Here we are presenting a selection of IoT and non-IoT systems.

#### 3.6.1. How Can We Obtain IoT Dataset?

We followed the method presented above to select the IoT dataset. In Step 1 in the process presented in Figure 1, we proceed to a manual selection based on a set of criteria. This allowed us to select systems that are relevant and mature to our research objectives to provide meaningful insights using those selection criteria:

- The repositories under IoT tags have different variants of syntax (Internet of Things, IoT, EIoT, IIoT, Industrial Internet of Things, Internet of Everything).

- Languages of the repository are supported by both used analysis tools (Java, JavaScript, C, C++, C#, Python).

- The number of stars is greater than 200 (ensuring that the system is well-rated and of good quality).

| Category | Metric | Definition | Formulas | Tool |
|---|---|---|---|---|
| Size | LOC | Counts the number of lines of source code in the system reflecting its size. In this work, LOC is calculated per file, and we sum all file values to have a value that represents the system. | LOC = Number of non-blank, non-comment lines in the code | Understand |
| | #Classes | Number of classes of each system | #Classes = Count of class declarations in the source code | Understand |
| | #Files | Number of files of each system | #Files = Total count of source code files in the project | Understand |
| Complexity | CC | Calculates the number of linearly independent pathways in system modules [19]. We computed the cumulative CC values by summing up the CC values of all classes within each application. | $CC(m) = E - N + 2P$<br>Where: $CC(m)$ is the cyclomatic complexity of control flow graph $m$, $E$ is the number of edges (transfers of control), $N$ is the number of nodes (a sequential group of statements containing only one transfer of control), and $P$ is the number of connected components. | Understand |
| | HV | Measures the software complexity used to assess the program size. The HV is used to measure the amount of code written. | $HV = N \cdot \log_2(n)$<br>Where: Total operators (N1) and total operands (N2), N: Program length calculated as N = N1 + N2, n: The vocabulary of your program is the sum of unique operators and unique operands. | Multimetric |
| | WMC | Measures the sum of the complexity of the methods in a class. This value is calculated per class; in this work, we sum up the WMC values of classes of each system. | $WMC = \sum_{i=1}^{N} CC_i$<br>Where: $CC_i$ McCabe's Cyclomatic Complexity of local method $i$, $N$ Total number of local methods in the class | Multimetric |
| Coupling | RFC | Measures the number of different methods and constructors that are called by a specific class. This value is calculated per class; in this work, we sum up the RFC values of classes. | $RFC = \text{Fan-In} + \text{Fan-Out}$ | Understand |
| | CBO | Assess the coupling between classes based on their usage. CBO metric measures the extent of coupling between two classes by examining the interactions between their methods and instances. The low value of CBO indicates low coupling [20]. This value is calculated per class; in this work, we sum up the CBO values of classes of the system. | $CBO = |C_{coup}|$<br>where : $C_{coup}$ set of classes | Understand |
| Cohesion | LCOM | Measures the count of separate sets formed by the local methods of a class, determined by their interaction with class variables [21]. High cohesion indicates good class subdivision [22]. We calculate the sum of LCOM values for each class in the system, and then we divide this sum by the number of systems to obtain the mean value of each. | $LCOM(C) = \frac{1}{a}\left(\frac{\sum_{j=1}^{a} \mu(Aj) - m}{1 - m}\right)$<br>Where: $a$ stands for the number of variables in a class $C$. $\mu(Aj)$ is the number of methods of $C$ accessing the variable $Aj$. $m$ stands for the number of methods in $C$. | Understand |
| Code Readability | CP | Quantifies the documentation level by measuring the proportion of code lines dedicated to comments. An appropriate documentation level is considered to be achieved when CP falls within the range of 20% to 30% [23]. CP is calculated per file; we sum all file values to have a value that represents the whole system. | $CP = \left(\frac{N_{comment}}{LOC}\right) \times 100\%$<br>Where: $N_{comment}$ is the total number of comments in the source code | Multimetric |
| Maintainability | MI | Measures the ease of maintaining a piece of software. Calculated based on metrics for a software system such as HV, CC, LOC, and the percentage of comment lines per module [24]. The higher the maintainability index, the easier it is to maintain the code. | $MI = 171 - 5.2\ln(HV) - 0.23CC$<br>$- 16.2\ln(LOC) + 50.0\sin(\sqrt{246 \cdot COM})$<br>Where: COM represents the percentage of comment lines per module. | Multimetric |

**Table 2**
Metrics used

- The number of forks is greater than 20.

- An active repository with the last push being at least 6 months ago (Date of last push greater than 04-2022).

- Mature repository created between 2012 and 2022.

### 3.6.2. How Can We Obtain non-IoT Dataset?

We used the same set of criteria used to select IoT systems for selecting the set of non-IoT systems. We build a query to choose a set of non-IoT systems with the same criteria as the selected IoT systems. This query was constructed to identify repositories on GitHub that met certain temporal, popularity, and technological criteria based on our IoT systems selection. The aim was to ensure that the selected repositories were recent, popular, actively maintained, and developed in languages relevant to our study, allowing us to analyse new, well-supported projects in the non-IoT domain.

Our query is:

> created: > 2012-10-01 created: < 2022-10-01 stars: > 200 stars: < 6500 forks: > 20 forks:< 20216 pushed: > 2022-04-01 language: C language: C# language: C++ language: Java language: JavaScript language: Python

**Stratification of the Query Output**

The execution of the query returns a large output. From this output, we select a representative dataset regarding the IoT dataset using stratification. Stratification is the process of dividing a dataset into homogeneous subgroups based on certain criteria. This approach allows for a more in-depth analysis within each subgroup and helps ensure that the datasets used for comparison (IoT and non-IoT systems) are as comparable as possible. To stratify the resulting dataset of non-IoT based on the criteria represented by the IoT dataset, we followed these steps.

1. We extracted pertinent details from the IoT repositories to serve as stratification criteria. We chose the composition of the programming language, the number of stars, and of forks. By considering these factors, we aim to create a subset of non-IoT systems that closely resembles the characteristics of the original set.

2. We create a mapping of criteria and repository names with each stratification criteria. For example, a dictionary maps programming languages to lists of repositories that use that language. We do the same thing with the stratification criteria.

3. We divide the non-IoT repositories into strata based on the relevant information.

4. For each subgroup, we select the repositories that most closely match the criteria represented by the IoT GitHub repositories. We use the Pareto principle [25] to select the top repositories in each subgroup.

5. We combine the selected repositories in each subgroup to form a representative subset of the data having the same number and characteristics of IoT systems.

### 3.6.3. How to Analyse and Verify the Two Sets?

In this step, we manually analyze the two selected sets to ensure they have an equal number of stars and forks, and use the same programming language, thus eliminating external factors that could affect the results. To examine data distribution, we perform statistical tests, including the Shapiro-Wilk test introduced by Hanusz *et al.* [26].

The Shapiro-Wilk test assesses the normality of data distribution. By checking if the data follows a normal distribution, it helps ensure the appropriateness of parametric statistical tests.

We compare the number of stars and forks in both datasets and use the non-parametric statistical Mann-Whitney U test [27] to determine if they are significantly different. The Mann-Whitney U test is a non-parametric test used to compare two independent groups when assumptions for parametric tests are not met (such as normal distribution). It is employed to determine if there are significant differences between the two datasets in terms of stars, forks, or other metrics. A U statistic value lower than $\approx 0.05$ indicates significant differences, while a higher U statistic suggests comparability between the datasets.

### 3.6.4. How to Identify Outliers?

Outliers are data points that are significantly different from the majority of the data [28]. They can have an impact on the results of statistical analyses. We remove outliers from our dataset to improve the accuracy of our results analysis.

We employed a meticulous approach to detect outliers within our dataset. Our methodology prioritized visual inspection, a recognized technique for outlier identification. Through visual representation, specifically by plotting the data, we aimed to pinpoint observations that deviated notably from the expected range. By systematically evaluating outliers and their potential impact on our analysis, we aimed to maintain the integrity and accuracy of our dataset. The removal of these influential outliers allowed for more reliable and precise statistical analyses moving forward.

The identification and removal of *outliers* is discussed in detail in Section 4.3.

### 3.7. How Did We Obtain non-IoT Best Practices?

Best practices refer to a set of recommended guidelines, approaches, methods, tools, or techniques that are considered optimal for reducing issues or enhancing the overall quality of the software. We approached the identification of these best practices systematically, initiating the process by formulating research queries tailored to each metric category and incorporating pertinent keywords. The query structure was designed as follows: **(X or Y) AND (software) AND (best practices)**, where X represents the category name, and Y relates to the specific practice associated with category reduction or improvement. For instance, in the context of the code size category, the query took the form of **((code size)**

OR (code reduction)) AND (software) AND (best practices). Subsequently, we executed these queries on Google Scholar, yielding varying numbers of papers for each category. Next, we employed a systematic selection process involving the filtration of the ten most highly cited articles that provided best practice insights for each category. We studied and extracted pertinent best practices from these articles. We assessed and categorised these extracted practices into three groups: *directly applicable*, *partially applicable with necessary adaptations*, or *not applicable to IoT*. This categorisation was guided by IoT-specific requirements, and the practices were subsequently prioritised based on their relevance and potential impact on the metric categories.

### 3.8. How Did We Ensure the Reproducibility of Our Selection and the Generalizability of Our Results?

We compared IoT and non-IoT systems, choosing the largest possible subset based on our criteria, believing it represents both types well. Our selection process considered various programming languages and system types to ensure diversity. Our dataset's relevance comes from methodically selecting diverse systems and using stringent criteria to ensure credibility and generalizability. By carefully choosing systems from GitHub and using stratification techniques, we ensured similarity and representativeness between IoT and non-IoT sets. Statistical analyses strengthened comparability and the credibility of our findings. While we could not cover every system, our careful process allows for reasonable generalizations to broader contexts for open-source systems available on GitHub. Also, we extended our systematic process to picking best practices systematically, guaranteeing reproducibility and validity.

## 4. Quantitative Analysis

To ensure the reproducibility of this work, we saved the code and the selection process in a replication package on Zenodo[1].

### 4.1. Output of Queries

We executed our queries, and we applied the process of selection presented in Section 3.6.1. The execution of our query of IoT systems yelled at 323 repositories. We removed 10 duplicate repositories. Next, based on manual verification of our selection criteria, we selected 94 repositories. We selected 94 comparable repositories using the stratification technique, presented in Section 3.6.2, of 1,972 repositories found when running the non-IoT search query.

### 4.2. Statistical Analysis of the two Datasets
#### 4.2.1. Nature of Distribution
The Shapiro-Wilk test results indicate that the calculated p-value is less than the significance level of 0.05; we have sufficient evidence to reject the null hypothesis. Meaning the number of stars and forks are not normally distributed.

#### 4.2.2. The Mann–Whitney U test Between the two datasets Regarding Stars and Forks Values
Results of the Mann-Whitney U test proved no significant difference between the distributions of stars and forks.

#### 4.2.3. Stars and Forks Distribution in the two Datasets
Figure 2 shows the scatter plot of the relationship between the stars in the function of forks. The dispersion of the data points in the scatter plot provides insight into the variability of the number of stars and forks. Most of the data points are tightly clustered together, which means that the number of stars and forks is similar across the two datasets.

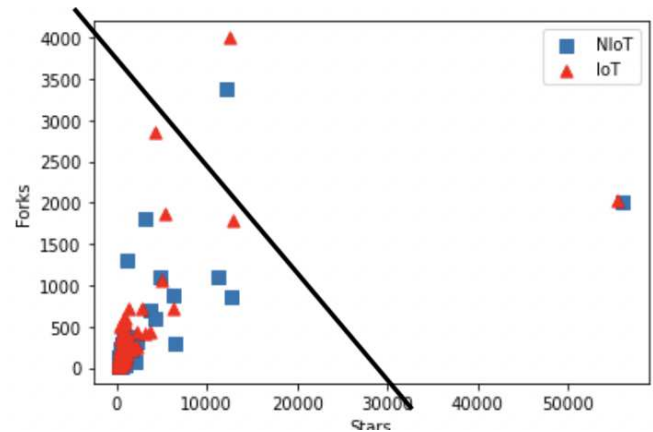

**Figure 2**: Stars and Forks Distribution in the Two Datasets

#### 4.2.4. Languages Distribution
We selected 14 Java, 11 C++, 16 JavaScript, 24 C, 25 Python, 4 C# systems.

### 4.3. Outliers
There are several methods for identifying and removing outliers. We use visual inspection to eliminate outliers by plotting the data and then identifying outliers. Any observations that fall outside the expected range are potential outliers. In Figure 2, visual inspection revealed outliers within our dataset. Notable instances included thingsboard/thingsboard and home-assistant/core for IoT systems and Apache/Druid and ansible/ansible for non-IoT systems.

To assess the influence of these outliers on our analysis, we associated them with top metric values (refer to Table 3). 'Apache/Druid' emerged as a system significantly affecting the results, thereby necessitating its removal from further analysis. Consequently, its corresponding IoT counterpart, 'thingsboard/thingsboard', was also excluded to maintain consistency in our dataset.

### 4.4. Top Values of Each Metric
We computed our metrics (CBO, RFC, LOC, WMC, CC, HV, MI, CP, LCOM and the number of Classes and Files). Table 4 presents the highest software metric values for IoT and non-IoT projects after removing outliers. IoT projects exhibit higher values in metrics like CC, RFC,

---

[1] https://zenodo.org/records/10564976

Comparison of Code Quality and Best Practices in IoT and non-IoT Software

|  | IoT | | non-IoT | |
|---|---|---|---|---|
|  | **Project Name** | **Value** | **Project Name** | **Value** |
| **RFC** | Samsung/TizenRT | 475185 | Apache/Druid | 191718 |
| **CBO** | eclipse-ditto/ditto | 83212 | Apache/Druid | 22423 |
| **CC** | espressif/esp-mqtt | 167 | quarnster/SublimeGDB | 339 |
| **HV** | Samsung/TizenRT | 7539487.09 | quarnster/SublimeGDB | 137964.39 |
| **MI** | eclipse-ditto/ditto | 368880.47 | Apache/Druid | 475261.25 |
| **LOC** | Samsung/TizenRT | 2009696 | Apache/Druid | 1003619 |
| **WMC** | project-chip/connectedhomeip | 83347 | Apache/Druid | 16057 |
| **LCOM** | rwaldron/johnny-five | 0.95 | rthenica/ffmpeg-kit | 0.93 |
| **CP** | ARMmbed/mbed-os | 70251.58 | Apache/Druid | 4705.6 |

**Table 3**
Top Before Deleting Outliers

|  | IoT | | non-IoT | |
|---|---|---|---|---|
|  | **Project Name** | **Value** | **Project Name** | **Value** |
| **RFC** | Samsung/TizenRT | 475185 | DarthFubuMVC/fubumvc | 64016 |
| **CBO** | eclipse-ditto/ditto | 83212 | DarthFubuMVC/fubumvc | 22423 |
| **CC** | flomesh-io/pipy | 75256385.40 | mgba-emu/mgba | 26505 |
| **HV** | greghesp/assistant-relay | 86796 | dachev/node-cld | 240076813.4 |
| **MI** | eclipse-ditto/ditto | 368880.47 | wmira/react-icons-kit | 811513 |
| **LOC** | Samsung/TizenRT | 2009696 | dachev/node-cld | 551449 |
| **WMC** | project-chip/connectedhomeip | 83347 | rthenica/ffmpeg-kit | 78624 |
| **LCOM** | rwaldron/johnny-five | 0.95 | rthenica/ffmpeg-kit | 0.93 |
| **CP** | ARMmbed/mbed-os | 70251.58 | UnknownShadow200/ClassiCube | 1413.95 |

**Table 4**
Top After Deleting Outliers

LOC, WMC, and CBO compared to non-IoT projects. We analysed these projects and found that IoT projects involve more complex hardware and software interactions driven by real-time processing needs. Non-IoT projects generally have higher HV and MI metric values (Table 4). This difference is due to non-IoT projects typically being less complex, influenced by distinct design and coding practices in non-IoT software development.

### 4.5. Statistical Computation on Metrics

Metrics reveal that IoT systems feature more extensive and complex code than non-IoT systems due to their hardware constraints, necessitating larger codebases. This highlights IoT's unique characteristics and the necessity to consider them in research and analysis.

Table 5 indicates that IoT systems exhibit greater interdependence than non-IoT systems. IoT's mean CBO is 4252.41, compared to non-IoT's 715.72, illustrating the higher interconnectedness in IoT systems. This interdependence makes IoT systems more challenging to maintain and modify, reflected in the MI values, with a median of 24854.31 for IoT and 36403.86 for non-IoT systems.

Furthermore, IoT systems have more classes and files compared to non-IoT systems. For example, IoT systems have a median of 116.5 classes, while non-IoT systems have 33. This is due to IoT's distributed nature, integrating advanced technologies and adapting to diverse device standards, which require a larger codebase.

## 5. Qualitative Analysis

We conducted an in-depth study of systems in pairs, one belonging to IoT and the other non-IoT, which are written in the same language. This exploration is to confirm the observations that we got from Table 5. We presented in Table 7 the systems that we analysed. As we obtained similar results for each programming language, we decided to showcase only the two Java-developed systems in the study of metrics.

### 5.1. Definition of Analysed Repositories

The two Java systems that we present are eclipse-ditto/ditto system for IoT and kymjs/CJFrameForAndroid system for non-IoT.

Eclipse-ditto/ditto [29] is a framework for managing digital twins. A digital twin is a virtual representation of a physical object or system, and Ditto provides a way to manage the data associated with these virtual representations. The Ditto repository is designed to support device connectivity, data modeling, access control, event processing and analytics.

ymjs/CJFrameForAndroid [30] an open-source repository for Android developers, providing a framework for building Android apps. The framework is designed to simplify and accelerate Android app development. The framework provides an architecture for building Android apps.

**Table 5**
Comparison Of Metrics Between the Two Datasets

| | | IoT | non-IoT |
|---|---|---|---|
| CBO | Median | **188** | 40 |
| | Mean | 4252.41 | 715.72 |
| | Mode | 0 | 0 |
| RFC | Median | **505** | 98 |
| | Mean | 22334.82 | 2956.8 |
| | Mode | 0 | 0 |
| LOC | Median | **4574** | 3713 |
| | Mean | 62960.13 | 31374.58 |
| | Mode | 0 | 0 |
| WMC | Median | **217** | 22 |
| | Mean | 4550.46 | 1667.51 |
| | Mode | 0 | 0 |
| CC | Median | **462032.02** | 465 |
| | Mean | 26953.31 | 2358.08 |
| | Mode | 2 | 0 |
| HV | Median | 2016 | **283436.35** |
| | Mean | 7283.76 | 5334.77 |
| | Mode | 0 | 441940.37 |
| MI | Median | 3645.21 | **4666.48** |
| | Mean | 24854.31 | 36403.86 |
| | Mode | 0 | 987.39 |
| LCOM | Median | **0.72** | 0.75 |
| | Mean | 0.70 | 0.74 |
| | Mode | 0.83 | 0.82 |
| CP | Median | **44.87** | 17.91 |
| | Mean | 2200.13 | 168.49 |
| | Mode | 0 | 0 |
| #Classes | Median | **116.5** | 33 |
| | Mean | 975.89 | 373.76 |
| | Mode | 0 | 0 |
| #Files | Median | **115** | 47.5 |
| | Mean | 984 | 356.24 |
| | Mode | 0 | 2 |

### 5.2. Comparison of Classes and Files

When comparing the most complex classes of IoT and non-IoT systems, we found that IoT systems have more highly complex classes. The most complex class in the IoT system has a complexity of 23, whereas, in the non-IoT system, it is only 7.

Furthermore, the largest file in the IoT system exceeds 2,500 lines of code, while the largest file in the non-IoT system surpasses 600 lines. Additionally, the largest function in the IoT system contains 290 lines of code, larger than the largest function in the non-IoT system, which has 65 lines. These differences arise from the distinctive nature of IoT systems, characterised by complex hardware compatibility, sensor integration, real-time data processing, diverse communication protocols, extensive data management, and customised business logic.

### 5.3. Comparison of Other Metrics

Table 8 shows difference between the measured metrics. By digging into the code, we found that high RFC in the eclipse-ditto/ditto repository is due to many tightly coupled classes. The project handles complex IoT data and protocols, with one class having 55 complex imports for gateway-service connections.

The code includes modules that handle the processing of data collected from various IoT sensors. There are functions to parse, filter, aggregate, and transform sensor readings. The repository also provides implementations of communication protocols commonly used in IoT systems.

We examined a system with extensive class inheritance, leading to high coupling and an elevated CBO value. The code also featured intricate logic and business rules, necessitating extensive interaction between objects, further increasing CBO and indicating low cohesion (as evident from LCOM value).

Eclipse-ditto/ditto exhibited a high CC metric due to its complex algorithms, workflow management, and diverse APIs for device interaction, device protocols, and communication patterns like AMQP, MQTT, and Apache Kafka. It showed a highly modular structure, contributing to a high WMC value.

Contrastingly, Kymjs/CJFrameForAndroid have low CBO by employing techniques like the Model-View-Presenter architecture for code decoupling. This framework also focused on minimising code volume while maintaining functionality.

The IoT project has a more complex and larger code amount with higher values in most of the metrics. IoT projects often involve integrating multiple hardware and software components and managing data communication between them, which can result in a higher number of classes, files, and lines of code, as presented in Table 5.

The non-IoT projects have a higher HV because they use more distinct operators and operands compared to the IoT project, as indicated in Table 5.

## 6. In-depth Code Analysis

In the previous sections, we assert that IoT projects involve more complex hardware and software interactions, contributing to higher code complexity. The goal of this section is to provide specific examples of each of the IoT systems analysed in depth to illustrate this complexity and how it manifests in IoT system codebases.

### 6.0.1. Java IoT System

As discussed above, the Java system **Eclipse-ditto/ditto** is highly complex.

To further analyze the extent of complexity, we selected one class named **ImplicitThingCreationMessageMapper**, which belongs to the package
`org.eclipse.ditto.connectivity.service.mapping`

This class is responsible for integrating new IoT devices or things into the Eclipse Ditto IoT platform, handling necessary configurations, policies, and message transformations

Table 6
Systems Analysed in depth

| Language | IoT | non-IoT |
|---|---|---|
| Java | eclipse-ditto/ditto | kymjs/CJFrameForAndroid |
| JavaScriprt | rwaldron/johnny-five | uuidjs/uuid |
| C | timmbogner/Farm-Data-Relay-System | unbit/spockfs |
| C++ | project-chip/connectedhomeip | zeek/zeek |
| C# | renode/renode | madskristensen/MiniBlog |
| Python | DT42/BerryNet | JohnHammond/msdt-follina |

Table 7
Selected Systems for Each Language

|  | IoT eclipse-ditto/ditto | Non-IoT kymjs/CJFrameFor-Android |
|---|---|---|
| #Stars | 414 | 412 |
| #Forks | 147 | 157 |
| #Classes | 7573 | 32 |
| #Files | 4917 | 25 |
| RFC | 130215 | 328 |
| CBO | 83212 | 254 |
| CC | 17056558.06 | 94 |
| HV | 13423 | 72196.47 |
| MI | 368880.47 | 2043 |
| LOC | 363467 | 2040 |
| WMC | 41499 | 238 |
| LCOM | 0.62 | 0.67 |
| CP | 5702.18 | 17.23 |

Table 8
Comparing Measured Metrics

for seamless device integration and management, which makes the codebase more complex.

The Listing 1 contains configurations and logic for message transformations. The code uses the lambda expression in Java in a nested manner to set configurations for the message mapper. Then, it is used with *method reference operator* to filter and map Header Configuration, increasing the complexity of the code.

```
1 thingTemplate = configuration.findProperty
2 (THING_TEMPLATE).orElseThrow(()->
    MessageMapperConfigurationInvalidException.
    newBuilder(THING_TEMPLATE).build());
3 commandHeaders=configuration.findProperty(
    COMMAND_HEADERS, JsonValue::isObject,JsonValue::
    asObject).filter(configuredHeaders->!
    configuredHeaders.isEmpty()).map(configuredHeaders->
    { /* ... */ }).orElseGet(()->DittoHeaders.newBuilder
    () /* ... */);
```

Listing 1: Code for Msg Transformation in Java-based IoT System

The code also deals with the creation and manipulation of IoT entities like Thing, Policy, ThingId, etc., as seen in methods like **getCreateThingSignal**, **createInlinePolicyJson**, and **validateThingEntityId**. We present some details of those methods in Listing 2. It contains multiple imports, several interfaces, and methods specific to the Eclipse Ditto IoT platform, which results in a complex codebase. As discussed in the previous example, the nested environment incorporates various expressions, contributing to the overall complexity of the code.

```
1 privateSignal<CreateThing>getCreateThingSignal(
    finalExternalMessage message, final String template
    ){}
2 Logic for creating a Thing based on the message and
    template
3 ... there are other similar methods handling IoT
    entities
```

Listing 2: Code for IoT Entities Creation and Manipulation in Java-based IoT System

### 6.0.2. JavaScript IoT System

**rwaldron/johnny-five** is a protocol-based IoT and Robotics programming framework. We analysed eg/nodeconf-radar.js file. The code is responsible for a radar-like display with simulated scanning motion and distance detection using hardware components and real-time data transmission to a web interface.

The code given below in Listing 3 is complex and need experts to understand it as it contains interactions and initialisation of hardware components, which requires understanding the pin configuration and range specifications for the servo motor and the Ping sensor from the 'johnny-five' library. The code contains some fictitious numbers, for example, "pin", which has a value of 12. Also, the range is defined between 0 to 170, but it is not clear what functions these numbers are performing.

The file contains also real-time data handling with WebSocket Connections (Socket.io), as showed in Listing 4. This code snippet used to set up Socket.io for real-time

```
1  var scanner = new five.Servo({ pin: 12, range: [0, 170]
     ↪ });
2  var ping = new five.Ping(7);
```

Listing 3: JavaScript Code for Interactions between Hardware Components in IoT System

communication. The use of Socket.io indicates the implementation of real-time data transmission, allowing communication between hardware and the web interface via WebSocket connections, which makes the codebase more complex compared to non-IoT systems. A complex codebase is less efficient in terms of resource utilisation and is difficult to reproduce.

```
1  var socket = require("socket.io");
2  var io = socket.listen(app);
```

Listing 4: JavaScript Code to Integrate Real-time Data Handling with WebSocket Connections in IoT Systems

Another proof of code complexity is its inclusion of concurrent operations, known as a factor contributing to complexity. As demonstrated by the code in Listing 5, which ensures the simultaneous management of servo scanning and Ping sensor data within event-based callbacks, presenting concurrent operations within the 'board.on("ready", function() ...)' callback.

```
1  this.loop(100, function(){}
2  Logic for scanning servo motion concurrently);
3  io.sockets.on("connection", function(socket){}
4  Event-driven handling of Ping sensor data while serving
     ↪ socket connections
5  ping.on("data",function(){}
6  Handling Ping sensor data concurrently);});
```

Listing 5: JavaScript Code to Demonstrate Simultaneous Operations in IoT Systems

### 6.0.3. C IoT System
The system **timmbogner/Farm-Data-Relay-System** uses ESP-NOW, LoRa, and other protocols to transport sensor data in remote areas without relying on WiFi. It is used for scenarios where there is a need for low-power. The code has high complexity due to several factors. The following code examples extracted from the file `fdrs_gateway_lora.h`.
The code in the file handles LoRa communication that involves multiple aspects such as frequency, spreading factor, power levels, ACK timeout, and retries, all of which contribute to configuring the radio for communication.

Listing 6 overviews a code example to define constants for LoRa configuration parameters like frequency, spreading factor, and transmission power. Configuring these parameters is crucial for effective communication but adds complexity due to their variety and specific values.

```
1  #define GLOBAL_LORA_FREQUENCY 915
2  Setting the LoRa frequency
3  #define GLOBAL_LORA_SF 12
4  Configuring spreading factor
5  #define GLOBAL_LORA_TXPWR 17
6  Setting LoRa transmission power
7  ...
8  ...
9  (other configuration constants)
```

Listing 6: C Code for Defining Configurations for LoRa

Also, functions to ensure LoRa Communication ultimately add to the complexity of the code. In Listing 7, a code snippet is provided where the 'transmitLoRa' function handles the construction and transmission of LoRa packets. It involves CRC calculation, packet assembly, and finally, transmitting the packet. This increases the complexity due to the detailed packet handling requirements.

```
1  crcResult transmitLoRa(uint16_t *destMac,DataReading
     ↪ *packet,uint8_t len){}
2  Logic for constructing and transmitting LoRa packets
     ↪ includes CRC calculation, packet construction, and
     ↪ transmission
```

Listing 7: C Code to Ensure LoRa Communication

In addition, asynchronous handling of LoRa transmission and reception introduces complexity, managing interruptions, flags, and different states for handling data transmission and reception simultaneously.

The code snippet of the 'setFlag' function, in Listing 8, manages interrupts and flags ('enableInterrupt', 'operationDone') to handle asynchronous communication. Complexity arises from managing interrupts and ensuring correct flag states for proper communication flow.

```
1  volatile bool enableInterrupt=true;
2  Flag to control interrupt
3  volatile bool operationDone=false;
4  Flag indicating packet sent/received
5  #if defined(ESP8266)||defined(ESP32)
6  ICACHE_RAM_ATTR
7  #endif
8  void setFlag(void)
9  Handling interrupt by setting operationDone flag
10 Enable/disable based on the enableInterrupt flag
```

Listing 8: C Code for Handling Asynchronous Communication in LoRa

### 6.0.4. C++ IoT System

**Project-chip/connectedhomeip** is a repository for a unified, open-source application-layer connectivity standard built to enable developers and device manufacturers to connect and build reliable and secure ecosystems and increase compatibility among connected home devices.

The code examined of the file `ContentAppPlatform.cpp` deals with dynamic endpoints and their associated attributes.

External callbacks for attribute read and write as shown in the Listing 9, through the method emberAfExternalAttributeReadCallback that handles attribute read operations, respectively, for dynamic endpoints. In the same file, there was a similar function emberAfExternalAttributeWriteCallback, which writes operations. The code checks whether the dynamic endpoint corresponds to a known content app. If found, it calls the app-specific handler; otherwise, it falls back to a generic handler. This demonstrates the complexity of managing different attribute operations based on dynamic endpoints and handling scenarios where the app is not available for a given endpoint which results in a complex codebase.

Managing access control for endpoints of IoT system introduces complexity. Listing 10 overviews code, which deals with setting and revoking permissions for various devices. It presents a function that manages access control by creating ACL entries and bindings for specific vendor and product IDs.

### 6.0.5. C# IoT System

We studied **renode/renode**, which is an open-source simulation and virtual development framework for complex IoT embedded systems.

We present a class named **ArduinoLoader** within the **Antmicro.Renode.Integrations** namespace from the file `ArduinoLoader.cs`

The class sets up USB devices, configurations, and functional descriptors. It involves configuring multiple USB interfaces, endpoints, and descriptors, which make the codebase complex, as presented in Listing 11.

```
1  EmberAfStatus emberAfExternalAttributeReadCallback
2  (EndpointId endpoint,ClusterId clusterId,const
       EmberAfAttributeMetadata* attributeMetadata,uint8_t*
       buffer,uint16_t maxReadLength){}
3  uint16_t endpointIndex=
       emberAfGetDynamicIndexFromEndpoint(endpoint);
4  ChipLogDetail(DeviceLayer,
       "emberAfExternalAttributeReadCallback endpoint%d",
       endpointIndex);
5  EmberAfStatus ret=EMBER_ZCL_STATUS_FAILURE;
6  ContentApp*app=ContentAppPlatform::GetInstance().
       GetContentApp(endpoint);
7  if(app!=nullptr){}
8  Handle attribute read based on dynamic endpoint
9  ret=app->HandleReadAttribute(clusterId,attributeMetadata
       ->attributeId,buffer,maxReadLength);
10 else
11 If the app is not found for the dynamic endpoint, use a
       generic handler
12 ret=AppPlatformExternalAttributeReadCallback(endpoint,
       rclusterId,attributeMetadata,buffer,maxReadLength);
13 return ret;
```

Listing 9: C++ Code for Determining Corresponding between Dynamic Endpoints

Also, as we are dealing with an IoT system, there is data transfer. The Decode method processes incoming data as shown in Listing 12. It iterates over the input data, interpreting ASCII characters. Depending on the character type, it appends nibbles to form numerical values. Switch statements handle special characters, indicating different types of commands. Code handles various cases, which makes the codebase complex.

### 6.0.6. Python IoT System

**DT42/BerryNet** is an AI/IoT system that connects independent components. Component types include but are not limited to AI engine, I/O processor, data processor (algorithm), or data collector. We studied the code in file `bnpipeline.py`, which defines classes related to a pipeline engine for processing data in an AI/IoT context.

The complexity of the studied codebase arises from its dynamic behaviour, extensive configuration options, communication handling, and the need to manage different modes and engines based on external messages. While these features provide flexibility to IoT, they also increase the overall complexity of the codebase.

In Listing 13, we present code ensuring dynamic engine switching between a real pipeline engine (PipelineEngine) and a dummy engine (PipelineDummyEngine) based on MQTT messages indicating the service mode (inference,

```
1   // Example of managing access control with ACLs
2   // and bindings
3   CHIP_ERROR ContentAppPlatform::ManageClientAccess(
        Messaging::ExchangeManager&exchangeMgr,
        SessionHandle&sessionHandle,uint16_t,targetVendorId,
        uint16_t targetProductId,NodeId localNodeId,std::
        vector<Binding::Structs::TargetStruct::Type>
        bindings,Controller::WriteResponseSuccessCallback
        successCb,Controller::WriteResponseFailureCallback
        failureCb){}
4   // Logic for managing ACLs and bindings
5   // Creation and handling of ACL entries and bindings
6   // based on vendor and product IDs
7   ...
8   ...
9   return CHIP_NO_ERROR;
```

Listing 10: Access Control with ACL's and Bindings using C++

```
1   USBEndpoint interruptEndpoint = null;
2   // ... (USB configuration)
3   USBCore = new USBDeviceCore(this,classCode:
        USBClassCode.CommunicationsCDCControl,maximalPacketSize:
        PacketSize.Size16,vendorId: 0x2341,productId:
        0x805a,deviceReleaseNumber:
        0x0100).WithConfiguration(configure: c =>
        c.WithEndpoint(Direction.DeviceToHost,
        EndpointTransferType.Interrupt, maximumPacketSize:
        0x08, interval: 0x0a, createdEndpoint: out
        interruptEndpoint))
4   // ... (configuring USB interfaces, endpoints, and
        descriptors)
5   // ...
```

Listing 11: C# Code for Setting Up USB Endpoints, Configurations and Functional Discriptions

```
1   private void Decode(byte[]d){}
2   this.Log(LogLevel.Noisy, "Decoding input:{0}", System.
        Text.ASCIIEncoding.ASCII.GetString(d));
3   uint value = 0;
4   uint savedValue = 0;
5   var command = Command.None;
6   for(var i=0;i<d.Length;i++){}
7   if(d[i]>='0'&&d[i]<='9'){}
8   AppendNibble(ref value,(byte)(d[i]-'0'));
9   else if(d[i] >= 'a' && d[i] <= 'f'){}
10  AppendNibble(ref value,(byte)(d[i]-'a'));
11  else if(d[i]>='A'&&d[i]<='F'){}
12  AppendNibble(ref value,(byte)(d[i]-'A'));
13  else{}
14  switch((char)d[i]){}
15  // ... (handling various cases)
16  }}}}
```

Listing 12: C# Code for Encoding and Decoding Incoming Commands

```
1   if mode=='inference':self.disable_engine=False
2   self.engine=PipelineEngine(...)
3   else:
4   self.disable_engine=True
5   self.engine=PipelineDummyEngine()
```

Listing 13: Python Code to Ensure Dynamic Engine Switching

```
1   self.comm.send('berrynet/engine/pipeline/result',
        tools.dump_json(generalized_result))
```

Listing 14: Python Code to Ensure Communication with an MQTT Broker

idle, or learning). This dynamic switching adds complexity to the code.

The code ensures communication with an MQTT broker, handling various topics and messages. This includes sending results, deploying newly retrained models, and switching between inference and non-inference modes. The use of MQTT for communication introduces complexity to the codebase. Listing 14 is an example of communication handling.

## 7. Practical Implications for IoT Development

We present a discussion on the implications of the observed values of the measured quality metrics and our in-depth analysis and their results on practical IoT development. We link those observations to real-world challenges, we provide implications for IoT developers and practitioners.

**Real-world challenge and observation 1**: IoT development involves intricate hardware-software interactions, intricate data communication, and complex algorithms. Metrics such as LOC, #Classes, #Files, CC, HV, and WMC highlight the extensive and complex nature of code in IoT systems. The in-depth example of code analysis also illustrates the difficulty of understanding IoT systems' code.

**Implication 1**: Developers engaged in IoT projects should cultivate specialized skills, such as encompassing a deep understanding of both hardware and software aspects and expertise in efficient data communication protocols to navigate challenges posed by hardware-software interactions, data communication intricacies and complex algorithms.

**Real-world challenge and Observation 2**: Higher interdependence between different modules within a system and low maintainability in IoT systems, reflected in metrics like RFC, CBO, and MI, pose challenges in making modifications and maintaining the codebase.

**Implication 2**: Emphasizing modular design and efficient code organisation allows for the encapsulation of functionality into distinct, manageable units, which is essential to effectively manage interdependence, maintainability, and extensive codebases in IoT systems.

**Real-world challenge and Observation 3**: The increased number of classes and files in IoT systems is driven by their distributed nature. The statistical computation on metrics reveals that IoT systems exhibit more extensive code compared to non-IoT systems. Metrics showcase a notable difference in the number of classes and files as presented in Table 5.

**Implication 3**: Developers engaged in IoT projects should prioritise the implementation of efficient organisational and structural practices, such as eliminating redundant code segments and optimizing function libraries. These practices involve code size reduction.

**Real-world challenge and Observation 4**: Understanding and maintaining code quality in evolving IoT projects. Understanding the trade-off between metrics like RFC, CBO, LCOM, CC, and WMC in IoT systems provides developers with actionable insights so they can make informed decisions and take specific actions in the development to enhance metrics values.

**Implication 4**: Continuous monitoring of code metrics, coupled with a willingness to adapt coding practices based on the results of these metrics, is essential for ensuring the quality of the IoT systems.

## 8. Results on Best Practices

Our study reveals differences between IoT and non-IoT systems. Therefore, we use these insights to enhance best practices from non-IoT literature to provide specific guidelines for addressing IoT-specific challenges discovered through our comparison, including high coupling, low cohesion, high complexity, low maintainability, code size reduction, and readability. We present best practices per category, shown in Table 9. Some of the found best practices, such as modularity and refactoring, can solve multiple problems, which is why we will find them repeated under different categories. We provide a more detailed version of Table 9, which includes additional information, in our replication package accessible on Zenodo[2].

### 8.1. Size

We observed that code size is bigger in IoT systems based on high values of the measured metric LOC and the increased number of classes and files in IoT systems. The execution of the query yielded 530,000 papers from which we selected the ten most highly cited papers. To solve the previously demonstrated size issues in Sections 4, 5, and 6, we found these best practices.

#### 8.1.1. Code Optimisation Techniques

Multiple studies introduced techniques to reduce the size of code [32, 33]. Most of these techniques could be used in IoT systems, such as loop unrolling, strength reduction by replacing costly operations with less resource-intensive alternatives, function inlining to minimise function call overhead, strength reduction of arrays, eliminating redundant computations, removing duplicated code, and optimising function libraries by selecting lightweight dependencies and unused code removal.

Static analysis tools can help to implement these techniques, such as Cppcheck, which is used in embedded systems and IoT development for C and C++ code [34]. Cppcheck might also detect opportunities to optimise loops or suggest better ways to handle iterations, indirectly impacting code size by reducing the number of instructions executed.

#### 8.1.2. Identification and Consolidation of Similar Functions

Reducing code size is possible through the identification and consolidation of similar functions.

One of the selected papers is the work of Edler *et al.* [33], which proposes a platform-independent code optimization technique to reduce code size by merging structurally similar functions. The Function Merging algorithm compares function signatures and control flow graphs to detect equivalence.

The platform-independent nature of the algorithm, operating at the intermediate representation level within a Low-Level Virtual Machine (LLVM), makes it adaptable to the diverse range of IoT devices with varying architectures by abstracting away hardware-specific details and allowing for the generation of code suitable for different target environments. The algorithm parameters, including minimum instruction count and similarity thresholds, contribute to its adaptability, ensuring that the merging process caters to the specific constraints of IoT environments.

#### 8.1.3. Use of Run-Time Decompression

Run-time decompression techniques provide code size reduction. This run-time decompression involves employing techniques such as dictionary-based software decompression and selective compression. Lefurgy *et al.* [35] proposed a dictionary-based software decompression, a software decompressor based on IBM CodePack, and a technique of

---

[2] https://zenodo.org/records/10564976

Comparison of Code Quality and Best Practices in IoT and non-IoT Software

| Category | Best Practices | Tools | Techniques | Availability | Aplaiability | Reason |
|---|---|---|---|---|---|---|
| Size | Code Optimisation Techniques | Cppcheck | Eliminate unused, duplicate code, and replaceable instances and detects opportunities to optimize loops | Yes | Yes | Cppcheck can be used to do code optimisation. |
| | Identification and consolidation of similar functions | NA | NA | No | No | Algorithms are available, but not any software. |
| | Use of Run-Time Decompression | IBM CodePack | NA | No | Yes | IBM's CodePack is not open source. |
| Complexity | Applying Refactoring | Eclipse, SonarQube, Checkstyle | Pulling up methods, extracting methods, and inlining methods | Yes | Yes | Tools are available and can be used. |
| | Applying Modularity | Docker, Virtual Box | Encapsulation and abstraction principles | Yes | Yes | Tools can be used. |
| | Use of Packaged Software Components | ThingSpeak, Microsoft Azure IoT Suite, Google Cloud IoT | NA | No | Yes | Available tools are not open source and will increase the code complexity of IoT systems. |
| Coupling and Cohesion | Application of design principles and patterns | NA | DI pattern for IoC, Patterns like Single Responsibility, and Singleton | Yes | Yes | Multiple studies have shown that design patterns can be applied to IoT systems. |
| | Applying Refactoring | Eclipse, SonarQube, Checkstyle | NA | Yes | Yes | Refactoring for IoT is possible by utilizing Eclipse. |
| | Applying Modularity | Docker, Virtual Box | Cluster Analysis Techniques[31] | Yes | Yes | Containerisation tools are available, like Docker. |
| Code Readability | Use of Textual Features | ESLint, Pylint, or Checkstyle | Use shorter lines of code and consistent indentation, comments, blank lines, meaningful and descriptive variables, etc. | Yes | Yes | Textual features must be integrated while implementing the system. |
| | Improve Code Entropy | NA | Enhance overall organisation, structure and variability of code | Yes | Yes | Developers must watch the value of entropy while implementing the system. |
| Maintainability | Use of Source code conventions | FindBugs1, Checkstyle2, and Jtest3 | NA | Yes | Yes | Best practice that must be integrated while implementing the system. |
| | Use of Model-Driven Architecture (MDA) | ThingML, Papyrus | NA | Yes | Yes | Appliacable when designing the system. |
| | Use of Design patterns | Eclipse | Factory Method, Singleton, and Decorator | Yes | Yes | Patterns can be used in the design phase. |
| | Applying Refactoring | Eclipse, SonarQube, Checkstyle | Encapsulation, limiting the length of code units to 15 lines of code, limiting the number of branch points per unit to 4, etc. | Yes | Yes | Refactoring for IoT is possible. |
| | Continuous Integration and Continuous Deployment (CI/CD) | Azure IoT Edge application, CircleCI, Jenkins as IoT CI/CD Manager | NA | Yes | Yes | To integrate in the development phase. |

**Table 9**
Best Practices for Various Code Categories

selective compression for controlling performance degradation due to decompression, using software-managed caches to support code decompression at the granularity of a cache line.

Techniques that we can adapt for IoT development from run-time decompression include selective decompression, dynamic decompression thresholds tailored to resource availability, and conditional decompression. However, whole-program decompression, real-time decompression of large codebases, and data compression techniques may be less practical for many IoT devices and components with limited memory and processing power.

### 8.2. Complexity

We observed that complexity is high in IoT systems based on high values of measured metrics such as CC and WMC. The query resulted in 366,000 papers, and we examined the ten most highly cited ones. To solve the previously demonstrated complexity issues in Sections 4, 5, and 6, we found these best practices.

#### 8.2.1. Applying Refactoring

Refactoring methods [36] provide an array of strategies for reducing complexity [37] .

Refactoring involves redistributing variables and methods across the class hierarchy to simplify the software system structure, with highlighted techniques such as pulling up, extracting, and inlining methods. Mayer Christian [38] underscores the importance of regular code refactoring in development for breaking down complex functions. While refactoring applies to IoT systems, it may introduce concurrency bugs and behaviour changes [39], requiring post-refactoring detection and evaluation for corrections.

For Java-based IoT systems, the refactoring process can be seamlessly executed in the Eclipse IDE, utilizing its integrated refactoring tools. Static code analysis tools like SonarQube or Checkstyle can identify potential areas for refactoring. Integrating them into the development pipeline for continuous static code analysis and improvement suggestions aids in reducing code complexity.

#### 8.2.2. Applying modularity

Modularity of the code is a prominent technique for complexity reduction, highlighted in Baldwin and Clark's theory of modularity [40] and in the work of Kearney *et al.* [41]. This technique emphasises the advantages of decomposing complex systems into smaller, manageable modules, a concept that finds resonance in IoT development. We believe that this principle can be applied to IoT systems by breaking down an IoT system into modular components. Containerisation tools like Docker are available to facilitate encapsulation and abstraction principles, which are techniques that contribute to the better modularity of the code. Docker uses containerisation to encapsulate applications and their dependencies, creating isolated environments. In IoT systems, we can create Docker containers for different components or services and package each component with its dependencies into a separate Docker image.

#### 8.2.3. Use of Packaged Software Components

Packaged software components are pre-built and ready-to-use software modules or frameworks that can be integrated into a larger software system. Their use is associated with decreased software complexity [42].

In IoT development, where this concept, like IoT platforms, is common, these findings bear significance. Examples of packaged software components include IoT platforms, which offer tools and services for building and managing IoT applications using tools such as ThingSpeak, Microsoft Azure IoT Suite, Google Cloud IoT, and IBM Watson IoT Platform.

### 8.3. Coupling and Cohesion

IoT systems have high coupling (high RFC and CBO) and low cohesion (low LCOM) compared to non-IoT systems. We found 26,400 papers and selected the top ten based on citations. From these ten papers, we extract best practices to solve the previously demonstrated coupling and cohesion issues in Sections 4, 5, and 6.

#### 8.3.1. Application of design principles and patterns

Walls and Breidenbach [43] showed that Dependency Injection (DI) achieves Inversion of Control (IoC), leading to reduced coupling and enhanced code cohesion.

In resource-constrained IoT environments, we do not have the luxury of using full-fledged DI frameworks. However, we believe that using lightweight DI Frameworks through a lightweight DI library such as TinyIoC or MicroDI, which are designed for embedded and IoT systems, is useful. These frameworks provide basic DI functionality without the overhead of larger frameworks.

Singleton and Factory [44, 45] ensure individual class responsibilities, enhancing cohesion and reducing coupling. For IoT, the application of these patterns is straightforward and facilitates the decoupling of modules, simplifying role separation and mitigating device heterogeneity [46]. When applying these patterns, there are minimal differences compared to non-IoT contexts that must be considered, as optimizing custom design pattern implementations for IoT systems operating in resource-constrained environments.

#### 8.3.2. Applying Refactoring

Same as for complexity Du Bois *et al.* [47] offered a guideline for refactoring to improve code coupling and cohesion. It is crucial to organise code with related functionality grouped and separate different concerns into distinct modules or classes [44]. Refactoring enhances coupling by reducing interconnections between modules through minimising method calls and shared variables [44].

We find that these refactoring principles apply to IoT systems based on our study of refactoring steps [47]. There are tools and IDE features available for developers to automatically identify and suggest refactorings for IoT systems [45], such as Eclipse, SonarQube, and Checkstyle.

### 8.3.3. Applying Modularity

Modularity is a well-known best practice to enhance coupling and cohesion. Utilising cluster analysis techniques can evaluate and improve modularisation [31]. In the IoT context, semantic categorisation can be employed to group IoT components based on their roles (e.g., sensors, actuators, controllers), and combining structural and semantic criteria enhances modularisation comprehensively.

To enhance modularity, cluster analysis techniques can be applied [31]. In the IoT system, this involves analysing relationships and dependencies between different components or modules. By identifying interrelationships among various IoT devices or components, we can create more cohesive and loosely coupled modules.

## 8.4. Code Readability

IoT systems exhibit higher code readability, as indicated by their higher CP values compared to non-IoT systems. The research query produced 66,700 papers, and we chose the ten most highly cited ones.

One significant challenge in readability studies is the complexity of experimentally substantiating what essentially constitutes a subjective perception. Obtaining measures of subjective perception is challenging, necessitating human studies and inherently involving variability. To derive useful measures, large-scale surveys that include multiple human raters and careful statistical analysis of inter-rater agreement are essential [48]. We report the found best practices that proved to be useful for improving code readability.

### 8.4.1. Use of Textual Features

Using simple textual features enhances code readability, emphasizing the importance of shorter lines, consistent indentation, and judicious use of comments [49, 50]. While comments may not uniformly indicate high readability, they directly communicate intent, making their use preferable. Blank lines, positively correlated with readability [49, 51]. Xiaoran *et al.* propose SEGMENT [52], a heuristic solution for automatic blank line insertion based on program structure and naming information.

Adapting SEGMENT's heuristics to IoT code by considering structural elements like event handlers, data processing, and communication tasks allows for inserting blank lines between logically related code segments, enhancing readability. Meaningful variable names and descriptive method names are important for clarity [51, 53]. In IoT development, employing clear and descriptive names for variables representing sensors, actuators, and data improves code readability, especially when methods interact with sensors or perform specific tasks.

To implement these techniques, manual code reviews focusing on the mentioned textual features or developing custom scripts tailored to IoT programming languages are viable options. Alternatively, existing static code analysis tools supporting readability metrics, such as ESLint, Pylint, or Checkstyle, can be adapted or extended to address the outlined requirements.

### 8.4.2. Improve Code Entropy

The concept of entropy measures the amount of information content in the source code. It is often viewed as the complexity, the degree of disorder, or the amount of information in a signal or data set. Entropy is calculated from the counts of terms (tokens or bytes) as well as the number of unique terms and bytes.

Posnett *et al.* [48] suggest that snippets with higher entropy are more readable. This implies that code with more varied elements (operators and operands) is easier to understand. When coding, developers must enhance the overall variability of code.

In IoT, we may be dealing with a variety of sensors, actuators, and communication protocols. Developers, when creating IoT systems, must consistently monitor the entropy value across diverse elements (operators and operands) in the code to enhance its overall entropy using static code analysis tools.

## 8.5. Maintainability

The maintainability of IoT systems is low compared to non-IoT systems; this is proved by the low value of MI, high code complexity, and high interdependence between different modules within a system. The research resulted in 55,300 papers, and we examined the ten most highly cited ones. To solve the previously found maintainability issues in Sections 4, 5, and 6, we found the following best practices.

### 8.5.1. Use of Source code conventions and Standards

Source code conventions and programming languages have evolved together, adhering to uniform conventions, such as naming conventions, inlined documentation, and syntactic structure, enhances code readability. Barry *et al.* outlined crucial code conventions for maintainability, particularly relevant to Java [54]. These conventions include recommendations for If, For, and Try statements, suggesting at most one additional nested statement, advocating the design of extensible classes without code in public methods, and more.

This conventions list holds wide applicability for IoT system code. Implementing these conventions can be facilitated by employing tools like FindBugs, Checkstyle, and Jtest.

### 8.5.2. Use of Model-Driven Architecture (MDA)

MDA involves expressing system requirements in a modelling language (e.g., UML) to generate a Platform Independent Model (PIM). This PIM is then transformed into a Platform Specific Model (PSM) for a particular technology and then into the actual code. MDA improves system maintenance by facilitating changes at the requirements level, automatically propagating them to affected modules [55].

In IoT development, MDA can be leveraged to create high-level models capturing system requirements and specifics like sensor integration, data processing, and communication protocols. Applying MDA in IoT ensures code generation based on these models, enhancing code maintainability and reducing errors [55].

### 8.5.3. Use of Design patterns

In addition to coupling and cohesion, design patterns positively impact code maintainability [56]. Jun *et al.* [57] empirically demonstrated that effective use of design patterns enhances software maintainability through an evaluation of a system without design patterns against its refined version after applying appropriate design patterns.

The use of design patterns in IoT systems is straightforward. For instance, the Factory Method pattern eases object creation without specifying concrete classes, facilitating the integration of new device types or functionalities in an IoT context. The Decorator pattern allows dynamic addition of responsibilities to objects, enabling flexible enhancement of IoT device capabilities without altering their core structure. Tools like Eclipse for Java systems can assist in implementing these patterns.

### 8.5.4. Applying Refactoring

Similar to complexity, coupling, and cohesion, refactoring techniques positively impact software maintainability [56] and reduce technical debt [58].

For C# code, Visser *et al.* provided guidelines for maintainability improvement through refactoring [59]. This includes limiting the length of code units (methods or constructors) to 15 lines, restricting the number of branch points per unit to 4 (splitting complex units into simpler ones), and balancing the relative size of top-level components.

Refactoring code in IoT solutions requires an understanding of the system architecture and its implications on data flow and communication protocols, facilitating code restructuring for improved maintainability without altering the external behaviour of IoT systems. While refactoring, we can implement encapsulation, which, as advocated by Anda and Bente [60], improves maintainability by hiding system details. In IoT, encapsulation involves concealing internal details of IoT devices and their communication protocols.

### 8.5.5. Continuous Integration and Continuous Deployment (CI/CD)

Implementing CI/CD pipelines to automate testing and deployment processes improves maintainability [58, 59].

For IoT systems, CI/CD facilitates rapid and reliable updates to IoT devices. However, there are specific considerations to consider before applying it to IoT. Such as creating realistic IoT device simulations for testing. Tools and frameworks like Eclipse Kapua, IoTivity, and IoT-LAB can simulate IoT device behaviour and interactions. Update mechanisms are essential for remotely deploying firmware updates to IoT devices. The whole CI/CD pipeline can be done through Azure IoT Edge application, CircleCI and Jenkins as IoT CI/CD Manager.

## 9. Threats To Validity

**Internal Validity:** Using a limited set of quality metrics may not comprehensively represent software systems. We selected various categories of commonly used metrics to address this limitation, ensuring a more holistic perspective on static code analysis.

We acknowledge that metrics alone are insufficient for a full-quality assessment. We need robust quality models in this task, which are based on quality metrics. Our work initiates this effort by measuring and comparing metrics, providing a foundation for future comprehensive models.

The choice of tools for measuring quality metrics may not align perfectly with the specific characteristics of IoT and non-IoT systems. To mitigate this concern, we employed two popular analysis tools instead of relying on a single tool, enhancing the accuracy of our results.

Our study focuses on various heterogeneous non-IoT systems, such as programming libraries, frameworks, databases, IDEs, games, scientific programs, etc. However, we acknowledge that more characteristics of these non-IoT projects could be integrated to select those systems, avoiding introducing biases or limitations due to inherent differences among these project types. We hope that future research will build upon our pioneering work, utilising more selection criteria to enhance the comprehensiveness and robustness of such analyses.

**External Validity:** Discrepancies in the experience levels of developers working on IoT and non-IoT systems can impact the differences in software quality. To address this potential bias, we conducted manual analyses to ensure the quality of selected systems. Additionally, we identified and addressed outliers and anomalies that could affect the validity of our results.

Our work may be susceptible to overlooking external factors influencing code metrics, such as environmental changes, specific hardware configurations, or external dependencies. We acknowledge that ignoring such factors might limit the accuracy and applicability of our findings.

**Conclusion Validity:** Comparing IoT and non-IoT systems is complex, as distinguishing between their code and systems is nuanced and complex. Despite these challenges, our work represents an initial step in this comparative analysis, laying the groundwork for future research.

We study a limited subset of open-source systems from Github that could threaten the generalizability and representativeness of the findings. These systems might not encompass the full breadth of diversity present in IoT and non-IoT systems, potentially limiting the applicability of the conclusions. Recognizing this limitation, we tried to mitigate this issue by considering the most extensive possible dataset available within the scope of our study to capture a diverse representation across different programming languages, frameworks, and project scales to achieve a more comprehensive understanding.

## 10. Conclusion and future work

The increasing prevalence of IoT systems emphasises the critical need to assess the quality of their source code, given their operation in resource-constrained environments and the complexity introduced by specific hardware requirements.

With IoT systems often serving in vital domains such as healthcare and infrastructure management, the impact of the code on functionality and reliability underscores the importance of careful code assessment throughout their development lifecycle.

This study addresses the existing gap in research on IoT systems software quality by conducting a comparative analysis with non-IoT systems software, acknowledging the unique challenges posed by IoT's limited resources and distributed architectures. The study and findings highlight key differences in metrics such as complexity, cohesion, code size, and maintainability, indicating that developing IoT systems demands tailored best practices.

Given these disparities, we systematically compiled a set of best practices commonly used in non-IoT systems and customised a list of best practices specifically designed for IoT system development to address these distinctions.

We systematically select and analyze 94 comparable IoT and non-IoT systems, providing comprehensive insights into their respective codebases. Our contributions include a method for choosing equivalent systems, computation and analysis of various metrics, an in-depth analysis of some IoT systems code, and a revisited list of software engineering best practices for IoT development, addressing observed challenges such as high complexity, low maintainability, and readability issues.

We acknowledge that recognising those metrics alone is insufficient for a complete quality assessment; this work sets the stage for future research, emphasising the implementation and evaluation of quality models to evaluate the quality of IoT systems. Overall, this study enhances our understanding of the software quality of IoT systems, providing insights for developing more resilient and efficient IoT systems across various domains.

Future research can further enrich our findings on software quality in IoT systems. Currently, the focus is on open-source systems available on GitHub; further research can extend the scope beyond that to include more systems and have a more diverse range of systems. While our study compared code quality between IoT and non-IoT systems via metrics, further investigations can build quality models and repeat the comparison process. Other aspects to explore in IoT systems include usability, security, and performance. Also, implementing identified best practices for non-IoT on IoT systems and evaluating their effects is necessary to address identified issues such as complexity, size, coupling, etc.

Comparison of Code Quality and Best Practices in IoT and non-IoT Software